\documentclass[aps,prb, twocolumn] {revtex4}
\usepackage{epsfig}

\begin{document}
\headsep 2.5cm

\title{Extra condition is necessary to have a unique cluster wave vectors set in the periodic cluster approximations. }
\author{Rostam Moradian$^{1,2,3}$}
\email{rmoradian@razi.ac.ir}
\affiliation{$^{1}$Physics Department, Faculty of Science, Razi
University, Kermanshah, Iran\\
$^{2}$Nano Science and Nano Technology Research Center, Razi University, Kermanshah, Iran\\
$^{3}$Computational Physical Science Research Laboratory, Department of Nano-Science, Institute for Studies in Theoretical Physics and Mathematics (IPM)
     ,P.O.Box 19395-5531, Tehran, Iran}

\begin{abstract}
We added an extra condition, original lattice symmetry of chosen cluster around cluster central site, to the cluster approximation methods with periodic boundary condition such as dynamical cluster approximation (DCA), effective medium approximation (EMSCA) and nonlocal coherent potential approximation (NLCPA). For each cluster size, this condition leads to a unique cluster wave vectors set in the first Brillouin zone (FBZ) where they preserve full symmetry of first Brillouin zone around ${\bf K=0}$. In this case whole cluster wave vectors are restricted to the FBZ and when number of sites in the cluster is equal to the whole lattice sites, these approximations recover original lattice symmetry in real and k-spaces.    
\end{abstract}
 \maketitle

\section { Introduction}  
  Different single site approximations such as coherent potential approximation (CPA)\cite{Soven67}, cluster approximations such as cluster CPA with open boundary condition, dynamical cluster approximation (DCA)\cite{Hettler98,Jarrell01,Jarrell01-2} and effective medium super-cell approximation (EMSCA)\cite{Moradian02, Moradian04, Moradian06} with periodic boundary condition are used to approximate self energy of interacting disordered systems. In the single site approximations the {\bf k} dependent of self energy is neglected while in the cluster approximations such as DCA\cite{Hettler98} for interacting and disordered systems, NLCPA\cite{Moradian02} for disordered systems and EMSCA\cite{Moradian06} in general for interacting disordered systems the cluster wave vector, {\bf K}, dependent of self energy are considered. The DCA originally constructed in the k-space by dividing the first Brillouin zone (FBZ) in to $N_{c}$ cells. The wave vectors at center of these cells are called cluster wave vectors and denoted by $\{{\bf K}_{1}, ...,{\bf K}_{N_{c}}\}$. They claimed these cluster wave vectors, ${\bf K}$, are corresponds to a real cluster with periodic boundary condition\cite{Hettler98}. But for some clusters there are many possible sets of cluster wave vectors, $\{{\bf K}\}$,\cite{Rowlands08} where some of cluster wave vectors are in the higher Brillouin zones. Recently to consider contributions of such different sets of cluster wave vectors, $\{{\bf K}\}$, the origin of FBZ fixed at ${\bf K}=(0,0,0)$ and a phase $\phi$ added to the cluster orthogonality condition,
\begin{equation}
 \frac{1}{N_{c}}\sum_{\bf K}e^{i{\bf K}.{\bf r}_{IJ}-i\phi}=\delta_{IJ}.
\label{eq:Row}
\end{equation}
 They claimed, when number of lattice sites $N_{c}\rightarrow\infty$ the phase, $\phi$, goes to zero and this relation converts to the following original lattice orthogonality condition, 
\begin{equation}
 \frac{1}{N}\sum_{\bf k}e^{i{\bf k}.{\bf r}_{ij}}=\delta_{ij}.
\label{eq:lattice ortho}
\end{equation}
 Their uncorrected assumptions and results are due to two factors, first their chosen cluster sizes haven't original lattice symmetry around cluster central site because this could lead to existence of different set of cluster wave vectors where some of wave vectors are in the higher BZs. Second, the Born von Karman periodic boundary condition, ${\bf k.L}=2\pi n$\cite{Ashcroft87}, imply that maximum number of cluster sites, $N_{c}$, should be the whole lattice sites $N$, not $N_{c}\rightarrow\infty$. So in this limit their defined $\phi$ is not going to zero, hence their orthogonality condition Eq.\ref{eq:Row} is not converting to the original lattice orthogonality condition Eq.\ref{eq:lattice ortho}. Since real lattice sites and allowed wave vectors in the FBZ have a one to one correspondence, hence for each cluster with $N_{c}$ sites, it must exist a unique set of $N_{c}$ cluster wave vectors ${\bf K}$ in the FBZ. Therefore we must add an extra condition to the periodicity of clusters with $N_{c}$ sites to have whole cluster wave vectors in the FBZ and just a unique set of super-cell wave vectors ${\bf K}$. Although periodicity condition for such real space clusters are considered\cite{Hettler98,Jarrell01,Jarrell01-2,Moradian02} but {\em symmetry of original lattice around central lattice site of clusters} are not considered. This is the weakness point of DCA and NLCPA where such symmetry are not considered. By considering both {\em periodicity} and {\em lattice symmetry conditions} for clusters, size of super-cells limit to those clusters that we can define the Wigner-Seitz cell (WSC) for their central sites and if we translate this WSC by cluster sites vectors set, $\{{\bf r}_{sc}\}$, where defined with respect to the central cluster site, they cover whole cluster with out overlapping. For a 2 dimensional square lattice (2d) this is illustrated in Figure \ref{figure:super-cell-symmetry}. This important condition lead to interesting result where for each cluster there is a unique super-cell wave vectors, $\{{\bf K}\}$, with full FBZ symmetry around ${\bf K}=(0,0,0)$. Here we reformulate the EMSCA in the real space to clarify this problem.

\section{Model and formalism}
We start our investigation by a general tight binding model for a disorder alloy system which is given by,       
\begin{eqnarray}
H&=&-\sum_{ij\sigma\sigma}t_{ij}c^{\dagger}_{i\sigma}c_{j\sigma}
\nonumber\\&+&\sum_{i\sigma} (\varepsilon_{i}-\mu)c^{\dagger}_{i\sigma}c_{i\sigma},
\label{eq:Hamiltonian}
\end{eqnarray}
where $c^{\dagger}_{i\sigma}$ ($c_{i\sigma}$) is the creation (annihilation) operator of an electron with spin $\sigma$ on lattice site $i$ and $\hat{n}_{i\sigma}=c^{\dagger}_{i\sigma}c_{i\sigma}$ is the number operator. $t^{\sigma\sigma}_{ij}$ are the random hopping integrals between $i$ and $j$ lattice sites with spin $\sigma$ respectively. $\mu$ is the chemical potential and $\varepsilon_{i}$ is the random on-site energy, where takes  $0$ with probability  $1-c$ for the host sites and $\delta$ with probability $c$ for impurity sites.      

The equation of motion for electrons corresponding to the above Hamiltonian, Eq.\ref{eq:Hamiltonian}, is given by, 
\begin{eqnarray} 
\sum_{l} \left(
       \begin{array}{c}
(E-\varepsilon_{i}+\mu)\delta_{il}-t_{il}\end{array}\right){ G}(l,j)=\delta_{ij}
\label{eq:equation of motion}
\end{eqnarray}
where $G(i,j)$ is the random single particle Green function. Eq.\ref{eq:equation of motion} could be rewrite as
\begin{equation}
 G(E;i,j)=G^{0}(E;i,j)+\sum_{l}G^{0}(E;i,l)\mbox{\boldmath$\varepsilon$}_{l} G(E;l,j).
\label{eq:random dyson equation}
\end{equation}
The Dyson equation corresponding to Eq.\ref{eq:equation of motion} for the exact averaged Green function, ${\bar G}(l,j)$, is\cite{ziman:79},
\begin{equation}
 {\bar G}(E;i,j)=G^{0}(E;i,j)+\sum_{ll'}G^{0}(E;i,l)\Sigma(E;l,l') G(E;l',j)
\label{eq:average dyson equation}
\end{equation}
where the real space self energy matrix, ${\bf \Sigma}$, is defined by,
\begin{equation}
\langle\mbox{\boldmath$\varepsilon$} {\bf G}\rangle={\bf \Sigma}\bar{{\bf G}}
\label{eq:self enrgy definition}
\end{equation}
By eliminate clean system Green function $G^{0}$ between Eqs.\ref{eq:average dyson equation} and \ref{eq:self enrgy definition}, the exact random Green function ${\bf G}$ could be expand in terms of exact average Green function $\bar{\bf G}$ and exact self energy ${\bf \Sigma}$ as,
\begin{equation}
 {\bf G}=\bar{\bf G}+\bar{\bf G}(\mbox{\boldmath$\varepsilon$}-{\bf \Sigma}){\bf G}.
\label{eq:random-average dyson equation}
\end{equation}
Note that although Eqs.\ref{eq:random dyson equation}, \ref{eq:average dyson equation} and \ref{eq:random-average dyson equation} are exact but no exact solutions for them. So they should be solved approximately. In the next section by adding {\em original lattice symmetry condition} to the chosen super-cell, we reformulate the effective medium super-cell approximation (EMSCA). In this new formalism we show for any super-cell size with {\em periodicity} and {\em original lattice symmetry} conditions not only there is a unique set of super-cell wave vectors ${\bf K}$ but also all super-cell wave vectors occur in the FBZ with its symmetry. 

\section{Reformulation of the effective medium super-cell approximation}
\begin{figure}
\centerline{\epsfig{file=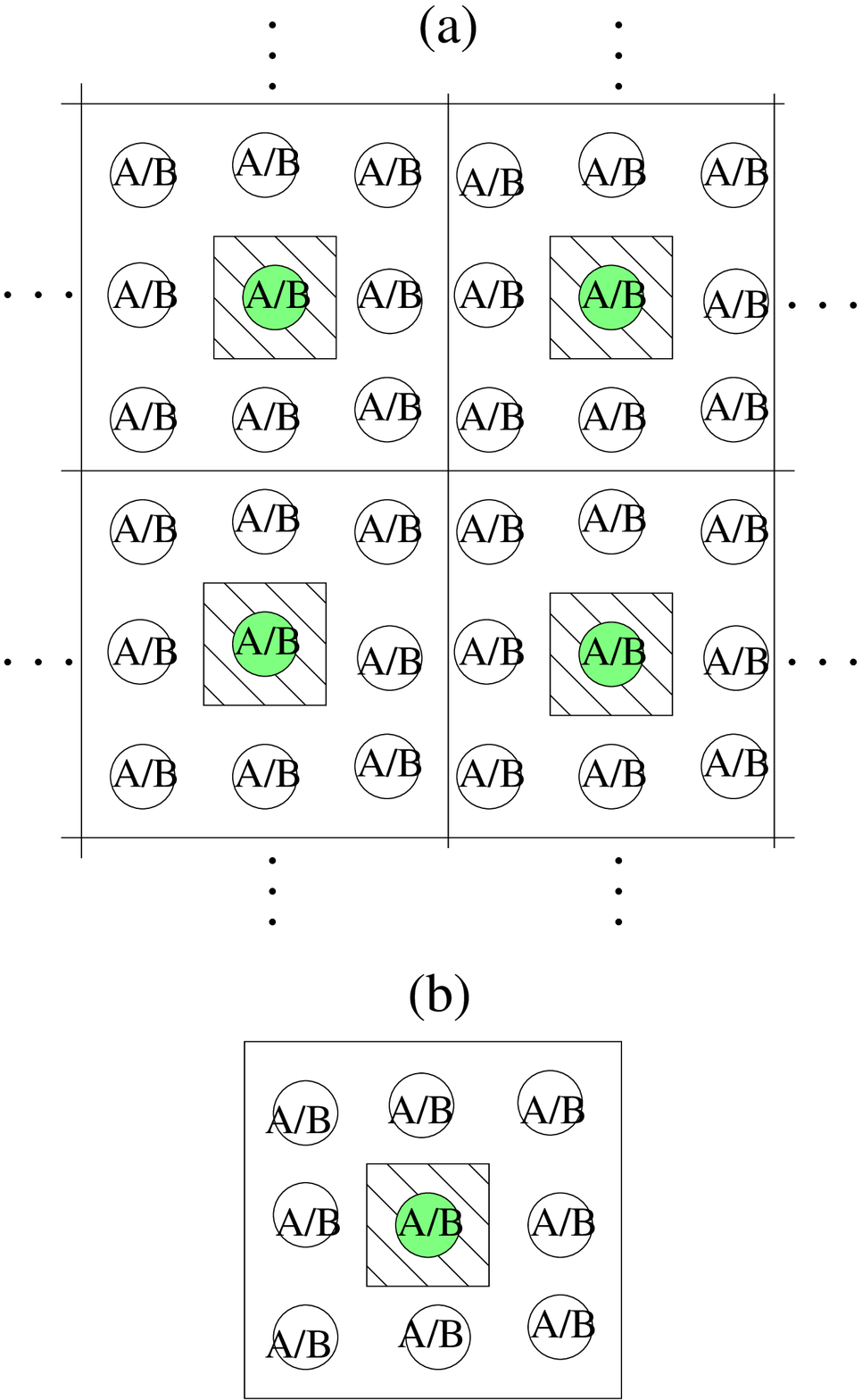 ,width=6.0cm,angle=0}}
\centerline{\epsfig{file=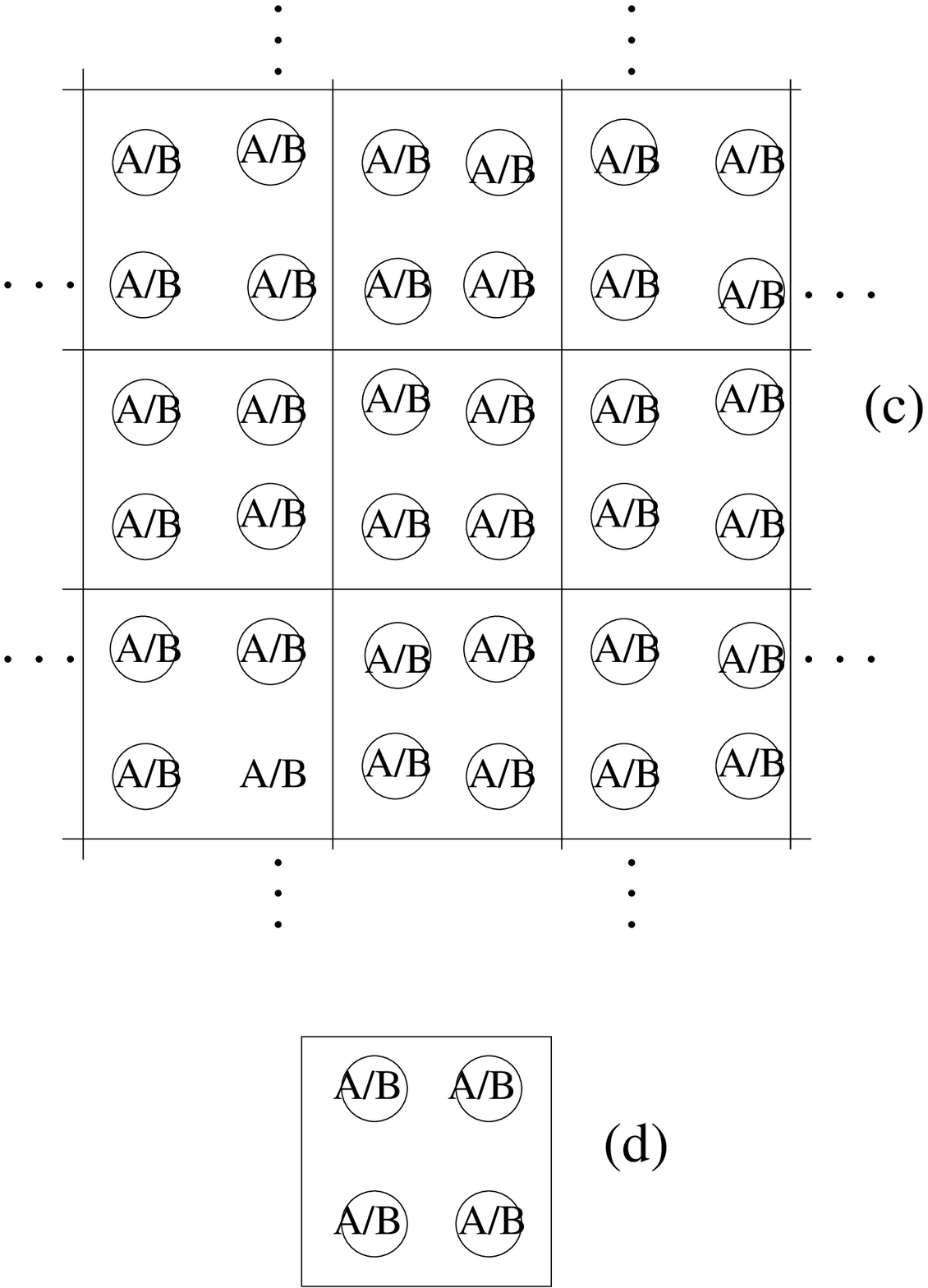 ,width=6.0cm,angle=0}}
\caption{(Color on line)(a) and (b) Show dividing a two dimension square real lattice in to similar super-cells of $N_{c}=9$ with original lattice symmetry around their own central sites. The green color sites illustrate central sites of super-cells and hatched squares show the Wigner-Seitz unite cell (WSUC) of each super-cell of this lattice. The WSUC and super-cells have whole original lattice symmetry around their origins. The super-cell vectors are ${\bf r}_{sc}=ma{\bf e}_{x} +na {\bf e}_{y}$ where $m,n=-1,0,1$ and {\em  a} is lattice constant. By translate each WSUC with respect to super-cell vectors ${\bf r}_{sc}=ma{\bf e}_{x} +na {\bf e}_{y}$, whole super-cell is cover with out overlapping. (c) and (d) show this 2d lattice divided in to four site clusters, $N_{c}=4$, where non of these clusters preserve original lattice symmetry around their own sites.  }   
 \label{figure:super-cell-symmetry} 
\end{figure}

\begin{figure}
\centerline{\epsfig{file=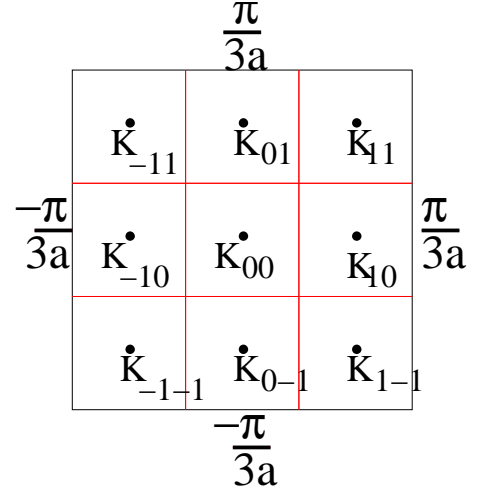 ,width=6.0cm,angle=0}}
\caption{(Color on line) Show FBZ of a 2d square lattice and its cells corresponding to super-cell with $N_{c}=9$. The unique set of super-cell wave vectors are ${\bf K}_{mn}=m\frac{\pi}{3a}{\bf e}_{x}+n\frac{\pi}{3a}{\bf e}_{y}$ where {\em m} and {\em n} take -1, 0 and 1. All these wave vectors are in the FBZ and the super-cell wave vectors, ${\bf K}$, have full FBZ symmetry around origin ${\bf K}={\bf 0}$}
 \label{figure:k-super-cell-symmetry} 
\end{figure}

\begin{figure}
\centerline{\epsfig{file=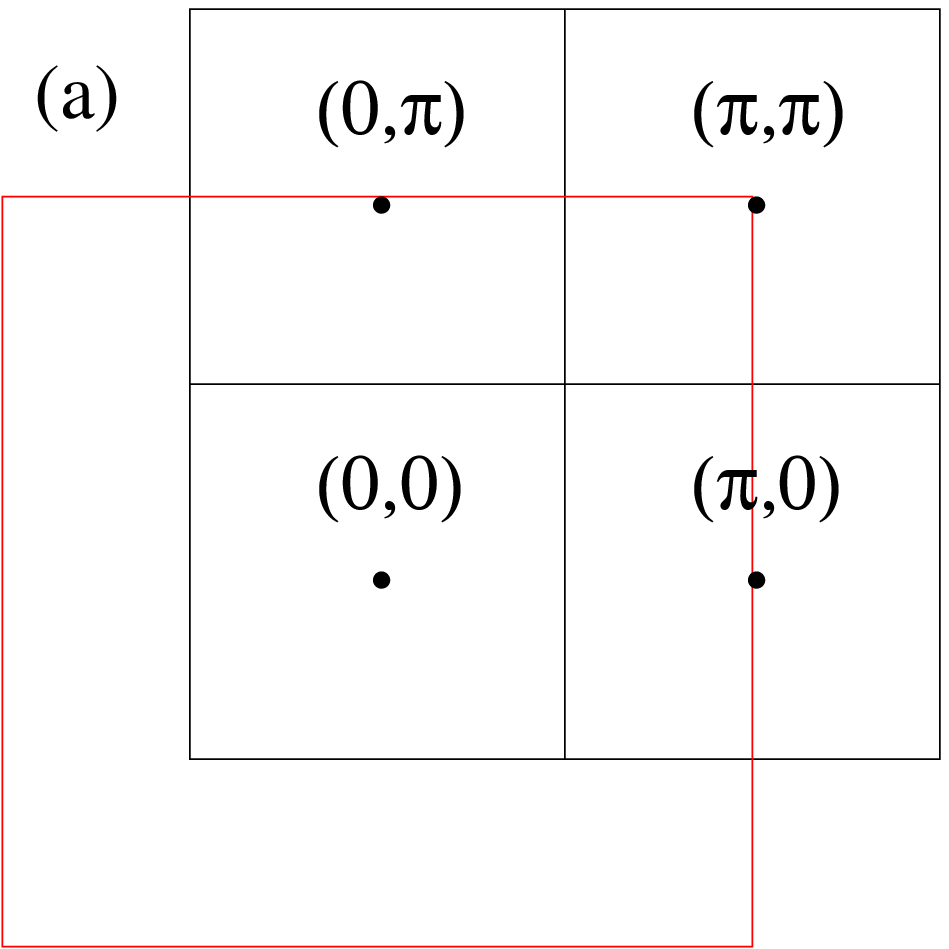 ,width=3.0cm,angle=0}\hspace{0.5cm}\epsfig{file=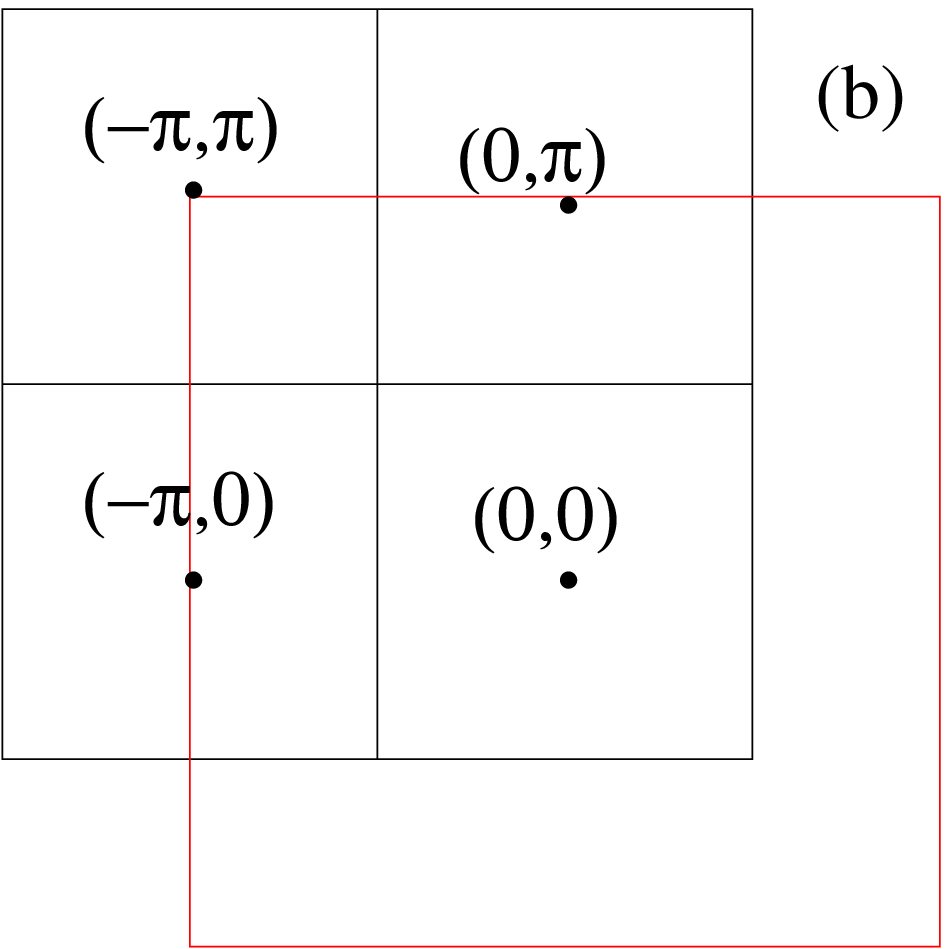 ,width=3.0cm,angle=0}}\vspace{0.5cm}
\centerline{\epsfig{file=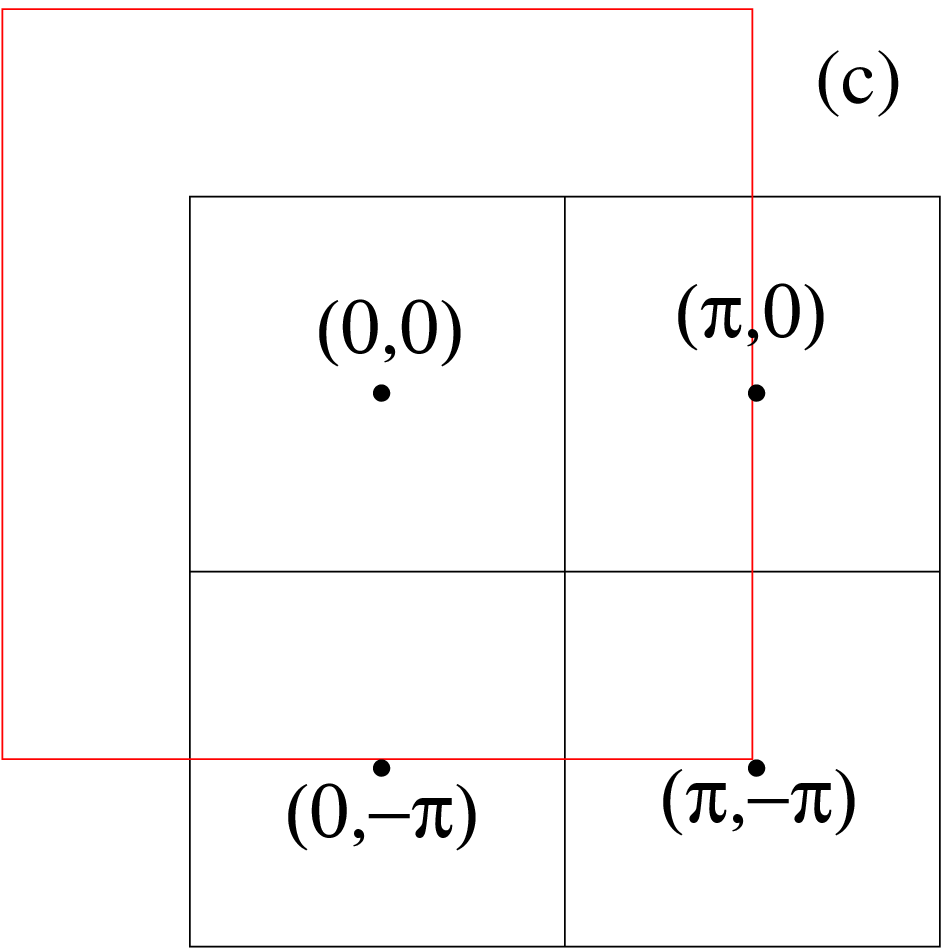 ,width=3.0cm,angle=0}\hspace{0.5cm}\epsfig{file=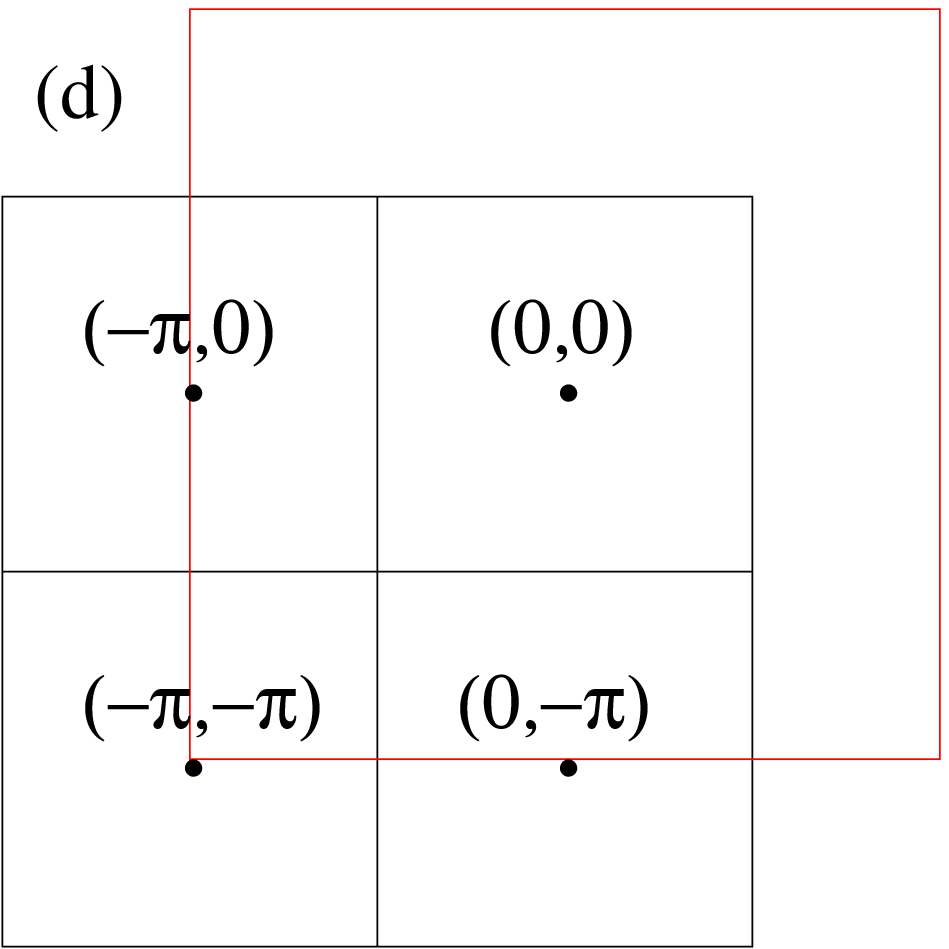 ,width=3.0cm,angle=0}}
\caption{(Color on line) Show four different super-cell wave vector sets $\{(0, 0), (0, \pi), (\pi, 0), (\pi,\pi)\}$, $\{(0, 0), (0, \pi), (-\pi, 0), (-\pi, \pi)\}$, $\{(0, 0), (\pi, 0), (0, -\pi), (\pi, -\pi)\}$ and $\{(0,0), (-\pi, 0), (-\pi,  -\pi), (0, -\pi)\}$ corresponding to four sites cluster approximation of Fig.\ref{figure:super-cell-symmetry} (c). The red square illustrate FBZ. Non of these cluster wave vector sets are in the FBZ and the super-cell wave vectors, ${\bf K}$, haven't FBZ symmetry around origin ${\bf K}={\bf 0}$. $a=1$ is the lattice constant.}
 \label{figure:k-super-cell-asymmetry} 
\end{figure}

In the EMSCA the real lattice is divided in to the similar super-cells such that each super-cell have whole original lattice symmetry around it's central site. This means that for each super-cell, if any one sit on it's central lattice site, exactly see whole lattice symmetry\cite{Moradian06}. Figure \ref{figure:super-cell-symmetry}(a) shows this approximation for a 2 dimension square lattice which is divided in to super-cells with nine sites, $N_{sc}=9$, the central site of each super-cell denoted by green color. Figure \ref{figure:super-cell-symmetry}(b) illustrate one of these super-cells. Each of these super-cells have original lattice symmetry around its own central site. The set of sites inside each super-cell denoted by $\{I\}$. Note that it is possible to divide this lattice in to super-cells with out lattice symmetry, for example in Figure \ref{figure:super-cell-symmetry}(c) and (d) this system divided in to square super-cells with four sites, $N_{sc}=4$, which super-cell hasn't original lattice symmetry around non of its own sites. Therefore this cluster can not give use a correct results. 
Since for each lattice vector there is a unique wave vector in the FBZ in the k-space, so for the set of super-cell vectors set $\{{\bf r}_{sc}\}$ it must exist a unique set of super-cell wave vectors in the k-space in the FBZ. To see lattice symmetry effect on the super-cell wave vectors, we reexamine the super-cell wave vectors corresponding to Fig.\ref{figure:super-cell-symmetry} (a) and (c). Fig.\ref{figure:k-super-cell-symmetry} shows unique super-cell wave vector sets corresponding to nine sites approximation, $N_{c}=9$, of  Fig.\ref{figure:super-cell-symmetry} (a).
 While for the four sites super-cell which is not preserved lattice symmetry, Fig.\ref{figure:super-cell-symmetry} (c), there are four different super-cell wave vector sets as illustrated in Fig.\ref{figure:k-super-cell-asymmetry}. The super-cell wave vectors of each set are not preserve FBZ symmetry. Hence we found just super-cells with original lattice symmetry around their own central sites preserve a one to one corresponding between real space super-cell vectors and super-cell wave vectors in the k-space. Therefore in the next part of this section we reformulate the EMSCA for super-cells with lattice symmetry.   

Since the super-cells in the real lattice are similar, intra super-cell multiple scattering kept and inter super-cell multiple scattering are neglected, hence the super-cell self energy, $\Sigma_{sc}(I,J)$, obey periodicity condition of the original lattice super-cells\cite{Moradian06},
\begin{equation}
\Sigma_{sc}({\bf r}_{IJ}+{{\bf L}^{mn}_{sc}})=\Sigma_{sc}({\bf r}_{IJ})
\label{eq:self enrgy periodicity}
\end{equation}
where ${\bf L}^{mn}_{sc}$ is a vector connects central lattice sites of {\em m}th  and {\em n}th super-cells,
\begin{equation}
{\bf L}^{mn}_{sc}=\sum_{i}(n_{i}-m_{i})N^{i}_{sc}{\bf a}_{i}
\label{eq:super-cell tranlation-vector}
\end{equation}
where $m_{i}$ and $n_{i}$ are integer numbers and $N^{i}_{sc}$ is number of lattice sites along the {\em i}th primitive vector ${\bf a}_{i}$. So $N^{i}_{sc}{\bf a}_{i}$ is length of super-cell along the ${\bf a}_{i}$ and for a 3-dimension system, number of sites in a super-cell is $N_{sc}=N^{1}_{sc}N^{2}_{sc}N^{3}_{sc}$. $m_{i}-n_{i}=l_{i}$ is an integer number. By applying Eqs.\ref{eq:self enrgy periodicity} and \ref{eq:super-cell tranlation-vector} to the following exact self energy relations,
 \begin{equation}
\Sigma(I,J)=\frac{1}{N}\sum_{\bf q}\Sigma({\bf q})e^{i{\bf q}.{\bf r}_{IJ}}
\label{eq:exact-q-self enrgy periodicity}
\end{equation}
we have
 \begin{equation}
\sum_{\bf q}\Sigma_{sc}({\bf q})e^{i{\bf q}.({\bf r}_{IJ}+{{\bf L}^{mn}_{sc}})}=\sum_{\bf q}\Sigma_{sc}({\bf q})e^{i{\bf q}.{\bf r}_{IJ}}.
\label{eq:q-self enrgy periodicity}
\end{equation}
where Eq.\ref{eq:q-self enrgy periodicity} imply that,
 \begin{equation}
e^{i{\bf q}.{{\bf L}^{mn}_{sc}}}=1
\label{eq:e-super-cell periodicity condition}
\end{equation}
hence,
 \begin{equation}
{\bf q}.{{\bf L}^{mn}_{sc}}=2\pi m'
\label{eq:super-cell periodicity condition}
\end{equation}
where {\em m'} is an integer number. Eq.\ref{eq:super-cell periodicity condition} for the nearest neighbour super-cells, $n_{i},m_{i}=0,1$, defines a unique set of $N_{sc}$ super-cell wave vectors in the first Brillouin zone (FBZ) which have FBZ symmetry around ${\bf K=0}$. To avoid confusing them with the lattice wave vectors {\bf k}, they are denoted by ${\bf K}=\{K_{n}\}$, 
 \begin{equation}
{\bf K}.{{\bf L}_{sc}}=2\pi m'
\label{eq:K-super-cell periodicity condition}
\end{equation}
where ${\bf L}_{sc}=\sum_{i}n_{i}N^{i}_{sc}{\bf a}_{i}$. From Eq.\ref{eq:K-super-cell periodicity condition} for each super-cell size we can obtain super-cell wave vector set $\{{\bf K}_{n}\}$. It should be emphasize that the super-cell wave vectors set, $\{{\bf K}_{n}\}$, around center of FBZ have original FBZ symmetry due to original lattice symmetry of super-cell. To reexamine this method we consider a limit case, when each super-cell extended to the whole lattice, that is $N^{i}_{sc}=N^{i}$, hence $N_{sc}=N$ and ${\bf L}_{sc}={\bf L}$. So the translation vector between nearest neighbour super-cells, where each super-cell are included whole lattice sites, are  
\begin{equation}
{\bf L}=\sum_{i}n_{i}N^{i}{\bf a}_{i}.
\label{eq:exact-super-cell tranlation-vector}
\end{equation}
Eq.\ref{eq:K-super-cell periodicity condition} by considering Eq.\ref{eq:exact-super-cell tranlation-vector} converts to the following periodicity condition
 \begin{equation}
{\bf K}_{(N_{sc}\rightarrow N)}.{\bf L}=2\pi m'.
\label{eq:super-cell Born van Karman periodicity condition}
\end{equation} 
The super-cell wave vectors set ${\bf K}_{(N_{sc}\rightarrow N)}$ defined by Eq.\ref{eq:super-cell Born van Karman periodicity condition} are exactly the lattice vectors ${\bf k}$ defined by the following periodicity Born von Karman condition, 
 \begin{equation}
{\bf k}.{\bf L}=2\pi m'
\label{eq:Born van Karman periodicity condition}
\end{equation} 
hence in this limit ${\bf K}_{(N_{sc}\rightarrow N)}={\bf k}$ and EMSCA become exact.

 In the next part of this section by applying EMSCA to the exact system we drive relation between real space and K-space super-cell self energies, average Green functions and the orthogonality relation. First we investigate self energies. In the EMSCA\cite{Moradian06} just self energy inside of each super-cell sites are non zero,
 \begin{equation}
\Sigma(i,j)=\delta_{iI}\delta_{jJ}\Sigma_{sc}(I,J)
\label{eq:super-cell approximation condition}
\end{equation} 
where {\em I} and {\em J} are restricted to the same super-cell. We apply Eq.\ref{eq:super-cell approximation condition} to the following exact relation to find relation between K-space and real space super-cell self energies, $\Sigma_{sc}({\bf K}_{n})$ and $\Sigma_{sc}(I,J)$,
\begin{equation}
\Sigma({\bf k})=\frac{1}{N}\sum_{ij}\Sigma(i,j)e^{i{\bf r}_{ij}.{\bf k}}
\label{eq:k-space real-space}.
\end{equation}
 Since number of super-cells in the whole lattice is $\frac{N}{N_{sc}}$ and in the super-cell approximation just self energies inside of each super-cell, $\Sigma_{sc}(I,J)$, are nonzero hence by using Eqs.\ref{eq:self enrgy periodicity} and \ref{eq:K-super-cell periodicity condition}, Eq.\ref{eq:k-space real-space} reduces to,
\begin{equation}
\Sigma_{sc}({\bf K}_{n})=\frac{1}{N_{sc}}\sum_{IJ}\Sigma_{sc}(I,J)e^{i{\bf r}_{IJ}.{\bf K}_{n}}
\label{eq:K-space real-space}.
\end{equation}
By converting Eq.\ref{eq:K-space real-space} to the real space we have
\begin{eqnarray}
\sum_{{\bf K}_{n}}\Sigma_{sc}({\bf K}_{n})e^{-i{\bf r}_{I^{'}J^{'}}.{\bf K}_{n}}&=& \\\nonumber\frac{1}{N_{sc}}\sum_{IJ}\Sigma_{sc}(I,J)\sum_{{\bf K}_{n}}e^{i({\bf r}_{IJ}-{\bf r}_{I'J'}).{\bf K}_{n}}
\label{eq:K2-space real-space}
\end{eqnarray}
where by consider the following super-cell orthogonality condition
 \begin{equation}
\frac{1}{N_{sc}}\sum_{{\bf K}_{n}}e^{i{\bf r}_{IJ}.{\bf K}_{n}}=\delta_{IJ}
\label{eq:super-K-space real-space}
\end{equation}
Eq.\ref{eq:K2-space real-space} reduces to,
\begin{equation}
\Sigma_{sc}(I,J)=\frac{1}{N_{sc}}\sum_{{\bf K}_{n}}\Sigma_{sc}({\bf K}_{n})e^{-i{\bf r}_{IJ}.{\bf K}_{n}}
\label{eq:super-cell-real-space}
\end{equation}
Eqs.\ref{eq:K-space real-space}, \ref{eq:super-K-space real-space} and \ref{eq:super-cell-real-space} are a set of equations that can convert real space $\Sigma_{sc}({\bf K}_{n})$ and K-space super-cell self energies $\Sigma_{sc}(I,J)$. Although relation between real space exact self energy,$\Sigma(i,j;E)$, and super-cell self energy, $\Sigma_{sc}(I,J;E)$, is known using Eq.\ref{eq:super-cell approximation condition} but in the {\bf K} and {\bf k}-spaces it should clarify. To find relation between exact self energy $\Sigma({\bf k})$ and super-cell self energy $\Sigma_{sc}({\bf K}_{n})$ we defining ${\bf k}={\bf K}_{n}+{\bf k}^{'}$, where ${\bf k}^{'}$ restricted to a sub-cell in the FBZ symmetry around super-cell vector ${\bf K}_{n}$. Hence in following exact self energy relation,
\begin{equation}
\Sigma({\bf r}_{IJ})=\frac{1}{N}\sum_{\bf k}\Sigma({\bf k})e^{i{\bf r}_{IJ}.{\bf k}}
\label{eq: real-space k-space}
\end{equation}
the summation over {\bf k} could be separated in to two parts, $\sum_{\bf k}=\sum_{{\bf K}_{n}}\sum_{{\bf k}^{'}}$, where {\bf k'} are wave vectors with respect to ${\bf K}_{n}$\cite{Moradian06} hence, 
\begin{equation}
\Sigma({\bf r}_{IJ})=\frac{1}{N}\sum_{{\bf K}_{n}}e^{i{\bf r}_{IJ}.{\bf K}_{n}}\sum_{{\bf k}^{'}}\Sigma({\bf K}_{n}+{\bf k}^{'})e^{i{\bf r}_{IJ}.{\bf k}^{'}}
\label{eq: rearange-real-space k-space}
\end{equation}
this exact self energy, $\Sigma(I,J)$, could be convert to the super-cell self energy, $\Sigma_{sc}(I,J)$, if we assume 
\begin{equation}
e^{i{\bf k}^{'}.{\bf r}_{IJ}}=1
\label{eq:ssuper-cell second condition} 
\end{equation}
and
\begin{equation}
\sum_{{\bf k}^{'}}\Sigma({\bf K}_{n}+{\bf k}^{'})=\frac{N}{N_{sc}}\Sigma_{sc}({\bf K}_{n}).
\label{eq:DCA-self energy}
\end{equation}
Hence we have 
\begin{equation}
\Sigma_{sc}({\bf r}_{IJ})=\frac{1}{N_{sc}}\sum_{{\bf K}_{n}}e^{i{\bf r}_{IJ}.{\bf K}_{n}}\Sigma_{sc}({\bf K}_{n})
\label{eq: real-space K-space}
\end{equation}
Note that similar to Eqs.\ref{eq:ssuper-cell second condition} and \ref{eq:DCA-self energy} are introduce in the DCA\cite{Hettler98,Jarrell01,Jarrell01-2}. Our next task is to find such relations for Green functions.

\begin{figure}
\centerline{\epsfig{file=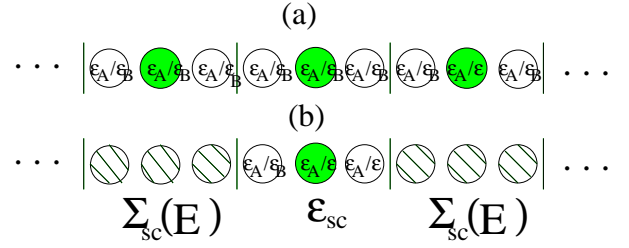, width=8.0cm,angle=0}}
\caption{(color on line) Show effective medium super-cell approximation method for a 1d alloy system. (a) Real 1d alloy system divided in to similar three sites super-cells, $N_{c}=3$ where each super-cell have full original lattice. The green color sites illustrate central sites of super-cells. By taking average over all super-cells except one where called {\em impurity super-cell} we have one impurity super-cell in an effective medium, where medium constructed from replacing super-cell impurities by super-cell self energies. (b) A super-cell with 3 sites, $N_{c}=3$, in an effective medium super-cell approximation. } 
 \label{figure:EMSCA-1d} 
\end{figure}

To find relation between average super-cell Green function ${\bar G}({\bf K}_{n})$ and exact average Green function ${\bar G}({\bf k})$,
\begin{equation} 
{\bar G}(I,J)=\frac{1}{N}\sum_{\bf k}{\bar G}({\bf k})e^{i{\bf r}_{IJ}.{\bf k}}
\label{eq:exact green function}
\end{equation}
where ${\bar G}({\bf k})$ is
\begin{equation}
{\bar G}({\bf k})=(G^{-1}_{0}({\bf k})-\Sigma({\bf k}))^{-1}
\label{eq:k-exact green function}
\end{equation}
and $G^{-1}_{0}({\bf k})$ is bare system Green function. Now we write each wave vector ${\bf k}$ in terms of its nearest super-cell wave vector ${\bf K}_{n}$ and inside cell wave vector ${\bf k}'$, ${\bf k}$=${\bf K}_{n}$+${\bf k}'$ hence $\sum_{\bf k}=\sum_{{\bf K}_{n}}\sum_{{\bf k}^{'}}$. By using these and Eq.\ref{eq:ssuper-cell second condition} the exact average Green function ${\bar G}(I,J)$ in terms of k-space Green function ${\bar G}({\bf k})$ is given by,
\begin{equation} 
{\bar G}(I,J)=\frac{1}{N}\sum_{{\bf K}_{n}}\sum_{\bf k'}{\bar G}({\bf k})e^{i{\bf r}_ {IJ}.{\bf K}_{n}}e^{i{\bf r}_{IJ}.{\bf k'}}
\label{eq:2exact green function}
\end{equation}
by using Eq.\ref{eq:ssuper-cell second condition} the exact average Green function, ${\bar G}(I,J)$, converts to super-cell Green function, ${\bar G}_{sc}(I,J)$,
\begin{equation} 
{\bar G}_{sc}(I,J)=\frac{1}{N}\sum_{{\bf K}_{n}}e^{i{\bf r}_{IJ}.{\bf K}_{n}}\sum_{\bf k'}{\bar G}({\bf K}_{n}+{\bf k'})
\label{eq:super-cell green function}
\end{equation}
where relation between K-space super-cell Green function, ${\bar G}({\bf K}_{n})$, and k-space exact average Green function, ${\bar G}({\bf k})$, is given by,
\begin{equation}
\sum_{\bf k'}{\bar G}({\bf K}_{n}+{\bf k'})=\frac{N}{N_{sc}}{\bar G}_{sc}({\bf K}_{n}).
\label{eq:exact-super-cell green function}
\end{equation}
similar to this equation derived in the DCA formalism\cite{Hettler98,Jarrell01,Jarrell01-2}. So relation between K-space and real space super-cell Green functions is given by,
\begin{equation} 
{\bar G}_{sc}(I,J)=\frac{1}{N_{sc}}\sum_{{\bf K}_{n}}e^{i{\bf r}_{IJ}.{\bf K}_{n}}{\bar G}_{sc}({\bf K}_{n})
\label{eq:r-K-super-cell green function}
\end{equation}
To complete our set of equations for calculation self energy and Green function,  we apply the EMSCA formalism to the self energy definition equation, Eq.\ref{eq:self enrgy definition}, where we take average over all super-cells except one, called {\em impurity super-cell}, hence the random potentials in all super-cells except {\em impurity super-cell} are replace by super-cell self energies\cite{Moradian06},
\begin{equation}
\mbox{\boldmath$\varepsilon$}{\bf\rightarrow} \left(
    \begin{array}{ccccc}
     & \vdots&\vdots& \vdots&\\
  ...&  {\bf \Sigma}_{sc} &{\bf \Sigma}_{sc}&{\bf \Sigma}_{sc}&...\\
  ...&  {\bf \Sigma}_{sc} &\mbox{\boldmath$\varepsilon$}_{sc}&{\bf \Sigma}_{sc}&...\\
  ...&  {\bf \Sigma}_{sc} &{\bf \Sigma}_{sc}&{\bf \Sigma}_{sc}&...\\
    &\vdots&\vdots&\vdots& \\
   \end{array}\right). 
\label{eq:EMSCA}
\end{equation}
Figure \ref{figure:EMSCA-1d} illustrate the EMSCA formalism for a 1d alloy system.
 Therefore Eq.\ref{eq:self enrgy definition} in the EMSCA reduces to,
\begin{equation}
\langle\mbox{\boldmath$\varepsilon$}_{sc} {\bf G}^{imp}_{sc}\rangle_{sc-imp}={\bf \Sigma}_{sc}\bar{{\bf G}}_{sc}
\label{eq:super-cell self enrgy definition}.
\end{equation}
By inserting Eq.\ref{eq:super-cell self enrgy definition} in to Eq.\ref{eq:random-average dyson equation} this equation reduces to the following relation for the super-cell impurity Green function for any arbitrary lattice sites {\em i} and {\em j}, 
\begin{eqnarray}
 & G^{imp}_{sc}&(i,j; E)=\bar{ G}_{sc}(i,j; E)+\nonumber\\&\sum_{LL'}&\bar{ G}_{sc}(i,L; E)(\varepsilon_{L}\delta_{LL'}- \Sigma_{sc}(L,L'; E))\times\nonumber\\&&   G^{imp}_{sc}(L',j; E).\nonumber\\
\label{eq:general super-cell random-green}
\end{eqnarray}
When {\em i} and {\em j} restrict to impurity super-cell sites, Eq.\ref{eq:general super-cell random-green} reduces to,
\begin{eqnarray}
 & G^{imp}_{sc}&(I,J; E)=\bar{ G}_{sc}(I,J; E)+\nonumber\\&\sum_{LL'}&\bar{ G}_{sc}(I,L; E)(\varepsilon_{L}\delta_{LL'}- \Sigma_{sc}(L,L'; E))\times\nonumber\\&&  G^{imp}_{sc}(L',J; E)\nonumber\\
\label{eq:super-cell random-green}
\end{eqnarray}
where
\begin{eqnarray}
 \langle G^{imp}_{sc}(I,J; E)\rangle=\bar{ G}_{sc}(I,J; E).
\label{eq:average-super-cell random-green}
\end{eqnarray}
Although Eq.\ref{eq:super-cell random-green} can be separated in to two equations by definition of {\em cavity Green function}\cite{Moradian06} but for disorder systems it is not necessary to defined such cavity Green function, although we must defined it in general for interacting disorder systems\cite{Moradian06}. 
Eqs.\ref{eq:exact-super-cell green function}, \ref{eq:r-K-super-cell green function}, \ref{eq:super-cell self enrgy definition}, \ref{eq:super-cell random-green} and \ref{eq:average-super-cell random-green} construct a complete set of equations to be solved self consistently to obtain super-cell self energy ,$\Sigma_{sc}(L,L'; E)$, and average super-cell Green function $\bar{ G}_{sc}(I,J; E)$. 
The algorithm for calculation of average Green function $\bar{ G}_{sc}(I,J; E)$ in the EMSCA is as follows

1- A guess for K-space self energies $\Sigma_{sc}({\bf K}_{\bf n})$ usually zero.
 
2- Calculate the super-cell average K-space Green functions, $\bar{G}({\bf K}_{\bf n})$, by inserting $\Sigma_{sc}({\bf K}_{\bf n})$ in $\bar{G}({\bf K}_{\bf n})=\sum_{\bf k'}(G^{-1}_{0}({{\bf K}_{n}}+{\bf k'})-\Sigma({\bf K}_{\bf n}))^{-1}$.

3- Fourier transform of K-space $\bar{G}_{sc}({\bf K}_{\bf n})$ to obtained real space Green function $\bar{ G}_{sc}(I,J; E)$.

4- Calculate $G^{imp}_{sc}(I,J; E)$ by inserting $\Sigma_{sc}(L,L'; E)$ and $\bar{ G}_{sc}(I,J; E)$ in to Eq.\ref{eq:super-cell random-green}.

5- Calculate average Green function $\bar{ G}_{sc}(I,J; E)$ by using Eq.\ref{eq:average-super-cell random-green} by taking average over all possible impurity configurations.

6- Calculate new self energies $\Sigma_{sc}(L,L'; E)$ by inserting obtained $G^{imp}_{sc}(I,J; E)$ and $\bar{ G}_{sc}(I,J; E)$ in Eq.\ref{eq:super-cell self enrgy definition}. 

\begin{figure}
\centerline{\epsfig{file=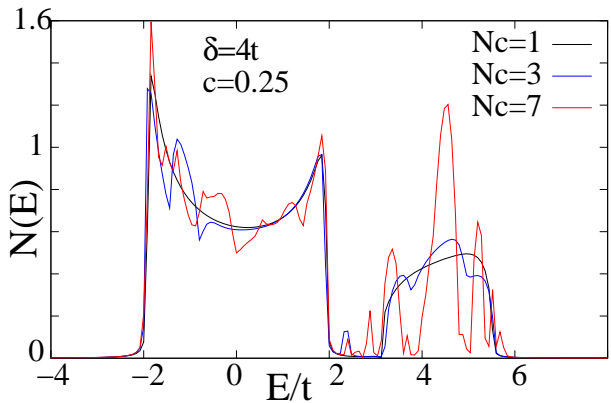 ,width=8.0cm,angle=0}}
\caption{(color on line) Show comparison of average density of states of a 1d alloy system for CPA, $N_{c}=1$, EMSCA with super-cell sizes, $N_{c}=3$ and $N_{c}=7$ where super-cell have original lattice symmetry. The strength length, $\delta=4t$, impurity concentration is $c=0.25$ and $\mu=0$. The difference between DOS are due to nonlocal corrections.} 
 \label{figure:k1-3-7-emsca-1d} 
\end{figure}

Now we apply our method to a 1 dimension (1d) alloy system. Since just super-cells with odd number of lattice sites preserve the lattice symmetry with respect to their middle site, we compared calculated average density of states for different super-cell size. Figure\ref{figure:k1-3-7-emsca-1d} illustrate average density of states for super-cells $N_{c}=1,\;3,\;7$. By increasing super-cell size more peaks in the average density of states. The corrections are due to nonlocal effects where neglected in the single site approximations, $N_{c}=1$.    

\section{Conclusion}
 We have added an extra condition to the cluster approximations such as Dynamical cluster approximation (DCA), non local coherent potential approximation (NLCPA) and effective medium approximation (EMSCA), where the chosen cluster must have original lattice symmetry with respect to its central site. This condition leads to a one to one correspondence between super-cell sites and super cell wave vectors in the first Brillouin zone hence a unique set of super-cell vectors for each cluster size. One of this vector is ${\bf K}=(0,0,0)$ and other super-cell vectors around this vector have full FBZ symmetry. While without this condition, for some cluster sizes, some of the cluster wave vectors locating in the higher Brillouin zones. This lead to two effects, first asymmetry of cluster wave vectors with respect to FBZ center hence when number of sites in the cluster goes to whole lattice sites the cluster wave vectors are not cover whole FBZ. Second, existence of many set of cluster wave vectors. In our formalism when number of cluster sites goes to number of lattice sites the super-cell vectors recover whole FBZ and EMSCA become exact.

\end{document}